\documentclass[10pt,prb,twocolumn]{revtex4-1}

\usepackage{graphicx}

\def\CHR{Cr$_2$O$_3$}

\begin{document}

\title{Microscopic origin of the structural phase transitions at the \CHR\ (0001) surface}

\author{A. L. Wysocki}
\email{alexwysocki2@gmail.com} \affiliation{Department of Physics and Astronomy and Nebraska Center for Materials and
Nanoscience, University of Nebraska-Lincoln, Lincoln, Nebraska 68588, USA}

\author{Siqi Shi}
\altaffiliation[Present address: ]{Department of Physics, Center for Optoelectronics Materials and Devices, Zhejiang
Sci-Tech University, Xiasha College Park, Hangzhou 310018, China}

\affiliation{Department of Physics and Astronomy and Nebraska Center for Materials and Nanoscience, University of
Nebraska-Lincoln, Lincoln, Nebraska 68588, USA}

\author{K. D. Belashchenko}
\affiliation{Department of Physics and Astronomy and Nebraska Center for Materials and Nanoscience, University of
Nebraska-Lincoln, Lincoln, Nebraska 68588, USA}

\date{\today}

\begin{abstract}
The surface of a Cr$_2$O$_3$ (0001) film epitaxially grown on Cr undergoes an unusual reentrant sequence of structural phase
transitions ($1\times1\to\sqrt3\times\sqrt3\to1\times1$). In order to understand the underlying microscopic mechanisms,
the structural and magnetic properties of the Cr$_2$O$_3$ (0001) surface are here studied using first-principles electronic
structure calculations. Two competing surface Cr sites are identified. The energetics of the surface is described by a
configurational Hamiltonian with parameters determined using total energy calculations for several surface supercells.
Effects of epitaxial strain and magnetic ordering on configurational interaction are also included. The thermodynamics
of the system is studied using Monte Carlo simulations. At zero strain the surface undergoes a
$1\times1\to\sqrt3\times\sqrt3$ ordering phase transition at $T_C \sim 165K$. Tensile epitaxial strain together with
antiferromagnetic ordering drive the system toward strong configurational frustration, suggesting the mechanism for the
disordering phase transition at lower temperatures.
\end{abstract}

\maketitle

\section{\label{sec:level1}Introduction}

Metal oxides demonstrate a variety of physical and chemical properties, sometimes in intriguing combinations. Apart
from being ubiquitous in nature, metal-oxide surfaces and interfaces find diverse technological applications and are
being explored for potential use in future electronic devices. In particular, surfaces of magnetoelectric
antiferromagnets such as \CHR\ possess an equilibrium surface magnetization, \cite{XiHe,Andreev,Kirill,NingWu} making
them suitable for use as active layers in electrically-switchable magnetic nanostructures. \cite{XiHe}

The \CHR\ (0001) surface has been a subject of many experimental
\cite{Bender,Rohr,Gloege,Maurice,Takano,Lubbe,Bikondoa} and theoretical
\cite{Bender,Rohr,Rehbein,Mejias,Cline,Wang,Rohrbach} studies, but its surface remains poorly understood. Low-energy
electron diffraction (LEED) experiments for a thin \CHR\ (0001) film grown on a Cr (110) single crystal revealed an
unusual reentrant structural phase transition,\cite{Bender} in which the surface structure changes from $1\times1$ to
$\sqrt3\times\sqrt3$ and back to $1\times1$ under cooling from room temperature to 150 K and then further down to 100
K. The origin of these phase transitions is not understood. While the high-temperature transition may, as suggested by
the LEED data, \cite{Bender} be a conventional order-disorder transition, the second one is unusual in that a more
symmetric phase appears at lower temperatures.

The situation is further complicated by the fact that, as shown by Takano \emph{et al.} \cite{Takano}, both phase transitions disappear for thicker \CHR\ films grown in a similar way.
This suggests that the epitaxial strain has an important effect on the surface energetics. Oxidation of the Cr $110$ surface was investigated by LEED and Auger spectroscopy, \cite{Ekelund} and it was found that
under growth conditions similar to those of Ref.\ \onlinecite{Bender} the thin \CHR\ (0001) film is subject to a tensile epitaxial strain of about 1.5\%.

In this paper we study the structure of the \CHR\ (0001) surface using first-principles electronic structure
calculations and Monte Carlo simulations. Our results suggest that the dynamics of the system is driven by the
occupation of two competing surface Cr sites. The system can be mapped to an Ising model on a two-dimensional hexagonal
lattice in external field. The first phase transition is clearly identified as a conventional ordering transition, and
the theoretical transition temperature is found to be in good agreement with experiment. Our calculations further
reveal a strong effect of tensile epitaxial strain, which parametrically drives the system towards configurational
frustration, particularly in combination with antiferromagnetic ordering. An explanation of the second phase transition
is offered based on these results.

The paper is organized as follows. In Section \ref{method} we describe the computational methods. Section
\ref{surf-struc} presents the results on the configurational and magnetic energetics of the \CHR\ (0001) surface, including the identification
of the competing surface Cr sites, the construction of the configurational Hamiltonian, the analysis of magnetic interactions, and the evaluation of the ground-state phase diagram. Section \ref{thermo} deals with configurational thermodynamics of the surface, and Section \ref{sec-exper} discusses the relation of the results to experiments. The electronic structure of the \CHR\ (0001) surface is presented in Section \ref{el-str}, and its magnetic properties in Section \ref{surf-mag}. Section \ref{summary} draws the conclusions.

\section{Computational methods}\label{method}

Electronic structure calculations were performed using the projector-augmented wave method \cite{Blochl} implemented in
the VASP code. \cite{Kresse,Kresse2} For the Cr $3d$ shell we employed the rotationally-invariant LSDA$+U$ method
\cite{Liechtenstein} with $U=4$ eV and $J=0.58$ eV. This method was preferred over GGA$+U$ adopted in Ref.\ \onlinecite{Rohrbach} due to its better description of the structural, electronic, and magnetic properties of bulk \CHR.\cite{Siqi} Different surface superstructures were modeled using
supercells representing symmetric slabs with eight atomic layers of O and 16 atomic layers of Cr stacked along the
(0001) direction. The periodically repeating slab is separated from its image by 1.5 nm of vacuum. We considered
$1\times1$, $1\times2$, $1\times3$, and $\sqrt3\times\sqrt3$ surface supercells (where $1\times1$ corresponds to the
hexagonal unit cell of bulk \CHR). The lateral dimensions of the unstrained supercell were fixed to the calculated
equilibrium bulk values;\cite{Siqi} for the strained case these values were used as a reference. Apart from these
constraints, the ionic positions were relaxed until the Hellmann-Feynman forces were converged to less than $0.01$
eV$/${\AA}. The plane-wave energy cutoff was fixed to $520$ eV and the Brillouin zone integration was performed using
$\Gamma$-centered Monkhorst-Pack grids. \cite{Monkhorst} For relaxation we used Gaussian smearing of $0.1$ eV and a
$k$-point mesh equivalent to or denser than $4\times4\times1$ for the $1\times1$ surface supercell. We checked the
convergence with respect to the number of $k$-points, the energy cutoff for the plane wave expansion, the size of the
vacuum region, and the thickness of the slab. These tests indicate that the total energies are generally converged to
within 1 meV. Density of states (DOS) calculations were performed using Gaussian smearing of $0.02$ eV and a $k$-point
mesh equivalent to or denser than $8\times8\times1$ for the $1\times1$ supercell.

The energy barriers for the thermally-activated jumping of Cr ions between the two competing surface sites were
calculated using the nudged elastic band method. \cite{neb} Seven images were inserted between the two energy minima,
and in each image the ions were relaxed so that forces perpendicular to the reaction path were smaller than 0.05
eV/\AA.

\section{Surface energetics} \label{surf-struc}

\CHR\ crystalizes in the corundum structure with the $R\bar3c$ space group. It can be viewed as a stacking of buckled
honeycomb Cr double layers along the (0001) direction with quasi-hexagonal closed-packed O layers in between, see Fig.\
\ref{slab}. The (0001) surface is polar, and simple electrostatic arguments suggest that non-stoichiometric
terminations by an O layer or by a Cr double layer should lead to divergent electrostatic potential in the bulk. On the
other hand, the surface can terminate in the middle of the buckled Cr layer so that only half of the Cr ions from this
layer remain on the surface. Although still polar, this termination is stoichiometric, and the electrostatic potential
in the bulk is not divergent. It can therefore be expected that this termination is energetically favorable. Indeed,
surface termination by a single Cr layer was consistent with LEED \cite{Rohr} and scanning tunneling microscope
\cite{Maurice} measurements of the \CHR\ (0001) surface in ultrahigh vacuum. Further, first principles calculations by
Rohrbach \emph{et al.} \cite{Rohrbach} based on the GGA$+U$ method have shown that this termination has the lowest
surface energy (compared to all others considered) over the entire range of oxygen chemical potential where \CHR\ is
stable. Note that earlier results based on the GGA method, which leads to grossly incorrect electronic and magnetic
properties,\cite{Siqi} were quite different.\cite{Wang}  In this work we only consider the \CHR\ (0001) surface terminated by a single layer of Cr.

\subsection{Surface sites}

\begin{figure}[htb]
\begin{center}
\includegraphics[width=0.45\textwidth]{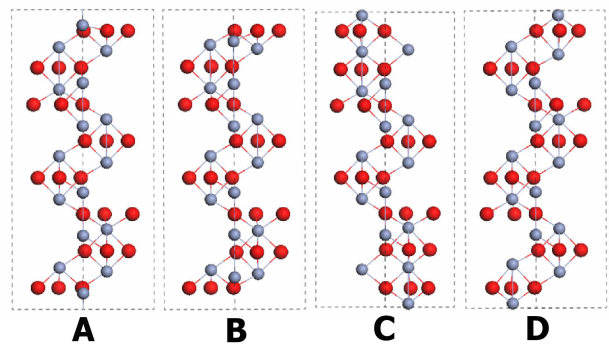}
\end{center}
\caption{Slab geometries for four considered $1\times1$ surface terminations. Gray and red spheres represent Cr and O
atoms, respectively.} \label{slab}
\end{figure}

The location of the Cr ions within the single Cr terminating layer has been debated. Within the double Cr layer there are
three possible octahedral sites, two of them being occupied in the bulk. They give rise to three nonequivalent surface
sites (A, C and D, see Fig.\ \ref{slab}) that surface Cr ions can occupy. Occupation of site A corresponds to the
continuation of the bulk structure. Further, as pointed out by Gloege \emph{et. al.}, \cite{Gloege} the surface Cr ion
can jump below the oxygen subsurface layer and occupy the empty octahedral site within the underlying Cr double layer.
This interstitial site is directly underneath the surface site A, and we denote it by B (see Fig.\ \ref{slab}).

In order to identify the energetically preferable sites, we therefore considered four $1\times1$ surface models
corresponding to the exclusive occupation of sites A, B, C, or D, respectively. In all cases a significant inward
relaxation was observed, as expected for a nominally polar surface. The relaxation data for models A and B are included
in Appendix \ref{App-A}. We define the surface energy as
\begin{equation}
E_s=\frac{1}{2}\left(E_{slab}-\frac{N_{slab}}{N_{bulk}}E_{bulk}\right)/N_s
\label{senergy}
\end{equation}
Here $E_{slab}$ is the ground state energy of the slab for the given surface model with magnetic structure
corresponding to bulk \CHR, $E_{bulk}$ is the ground state energy per unit cell of bulk \CHR, $N_{slab}$ and $N_{bulk}$
are the numbers of atoms in the slab and in the bulk unit cell, and $N_s$ is the number of surface Cr atoms on one side
of the slab. The surface energies for the four $1\times1$ surface terminations are given in Table \ref{tab1}. The
surface energy is the lowest when site A is occupied. Occupation of sites C and D leads to much higher surface
energies, and we therefore do not consider their occupation in the subsequent analysis. On the other hand, the surface energy of model B is only slightly higher than that of model A. Thus, sites A and B can both be partially occupied, which can lead to non-trivial ordered terminations and phase transitions; these issues are addressed in the following subsections.

\begin{table}
\caption{Surface energies of $1\times1$ surface models with different surface sites occupied.}
\begin{tabular}{|c|c|c|c|c|}
\hline
 & A  & B  & C  & D  \\
\hline
$E_s$, eV & 2.909 & 3.077 & 5.002 & 5.847 \\
\hline
\end{tabular}
\label{tab1}
\end{table}

The identification of sites A and B as the most favorable agrees with LEED measurements and molecular dynamics simulations of Ref.\ \onlinecite{Rohr}, as well as with surface X-ray diffraction data, \cite{Gloege}  but recent LEED \cite{Lubbe} and surface X-ray diffraction (SXRD) \cite{Bikondoa} studies have questioned the single Cr layer surface termination and reached different conclusions. In Ref.\ \onlinecite{Lubbe} a non-stoichiometric surface with a partial occupation of four Cr layers near the surface was obtained, but the best-fit $R$-factor $R_p=0.48$ was poor, as noted by the authors. In Ref.\ \onlinecite{Bikondoa} the best fit for the SXRD measurements was obtained for a surface terminated with a partially occupied double Cr layer (sites A and C) and one more partially occupied Cr layer below that. Partial occupancy of site C is difficult to reconcile with the very high surface energy  of surface model C (2 eV per Cr site higher compared to model A), although this site could, in principle, be stabilized by intersite interactions or by depletion of Cr atoms in the subsurface Cr layers. Since the configurational models required to explore such unconventional terminations would be very complicated, we did not attempt to consider them. We also note that the site occupations must be integer in the ground state. Further analysis may be required as more experimental evidence becomes available.

While the spin-orbit coupling in bulk \CHR\ is small, \cite{Siqi} the reduced coordination could make it more important at the surface. To estimate its role, we calculated the energy difference between surface models A and B in the presence of spin-orbit coupling (taking the structures relaxed without it). It was found that spin-orbit coupling changes this energy difference by $0.4$ meV. This energy is small compared to all important structural and exchange interaction parameters, and therefore spin-orbit coupling was neglected in all subsequent calculations.

\subsection{Configurational interaction at the surface}\label{config}

Surface A sites form a two-dimensional hexagonal lattice, and there is a B site directly underneath every A site. Based
on the surface energies calculated in the previous subsection, we assume that at every hexagonal lattice site the Cr
atom occupies either site A or site B. Therefore, we can introduce an occupation number $n_i$, where $i$ denotes a 2D
hexagonal lattice site, such that $n_i$ is equal to 1 if site B is occupied and 0 if site A is occupied. The following
configurational Hamiltonian can therefore be introduced:
\begin{equation}
\mathcal{H}= V_{int}(\{n_{i}\})+ h\sum_{i}n_{i}
\label{Hamiltonian}
\end{equation}
The first term includes the configurational interaction between surface Cr ions, and the second term takes into account
that sites A and B are inequivalent. Since the total number of A sites is not conserved, this Hamiltonian is isomorphic
to an interacting Ising model on a 2D hexagonal lattice in external magnetic field.

\begin{figure}[htb]
\begin{center}
\includegraphics[width=0.3\textwidth]{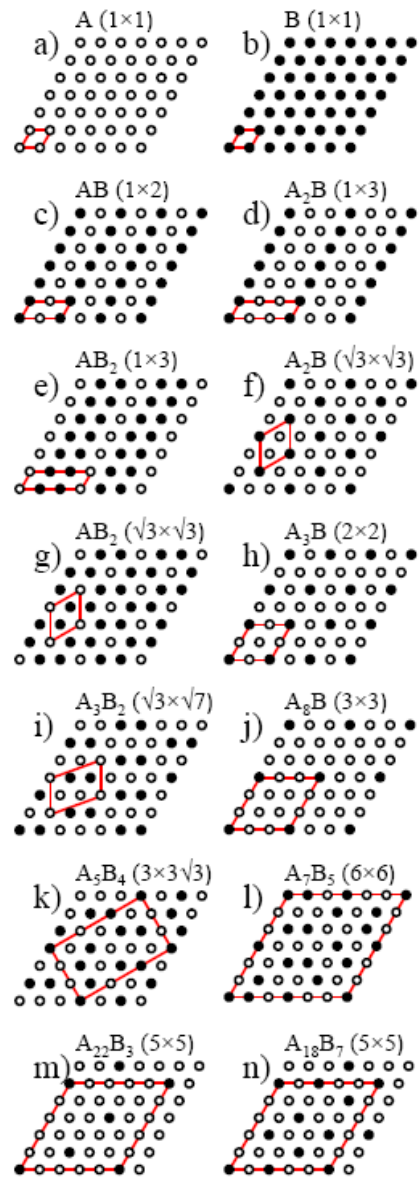}
\end{center}
\caption{Surface configurations mentioned in the text. Configurations (a)-(g) were used to fit the parameters of  the configurational Hamiltonian. Configurations (a), (c), (f) and (h)-(n) have been identified as possible ground states.} \label{structures}
\end{figure}

The introduction of the 2D hexagonal lattice is based on the spatial arrangement of A sites on the surface. Note,
however, that the true symmetry of the (disordered) \CHR\ surface is lower: apart from the translations, there are only $C_3$ axes passing through the Cr sites. To take this difference into account, one can formally assign a direction to each bond on the 2D hexagonal lattice. The directions of the six nearest-neighbor bonds should be made alternating (i.\
e.\ three incoming and three outgoing bonds). The directionality of the bonds can be reflected in the interaction term
in the Hamiltonian (\ref{Hamiltonian}). For example, the pair interaction parameter may be different for a bond
pointing from site A to site B and for a bond pointing from site B to site A. However, we are mainly interested in the
total energies of different configurations $\{n_{i}\}$ which are not strongly affected by the directionality of the
bonds. In fact, it can be shown that for pairwise interaction of any range the total energies do not depend on whether
the bond directionality is included or not. (This is because the total numbers of A$\to$B and B$\to$A bonds are always
equal in all coordination spheres.) Even when many-body interactions are present, the contribution of the bond
directionally to the total energy is zero for most ordered configurations. In particular, among all configurations
shown in Fig.\ \ref{structures} only A$_7$B$_5$ ($6\times6$) has a nonzero contribution, but thus structure is not
important for any of the following. Moreover, as we will show below, the surface structure is governed by
non-directional electrostatic interactions. We therefore do not introduce any non-directional terms in the
Hamiltonian, which makes the assignment of bond directions superfluous.

In order to proceed, we use the cluster expansion approach, \cite{Connolly,Sanchez,Fontaine,Zunger} which is widely used in the studies of bulk
alloy thermodynamics. Specifically, we need to adopt some particular representation of $V_{int}(\{n_{i}\})$ and fit it
to the calculated total energies of different ordered configurations $\{n_{i}\}$. However, due to large size of the
system the calculations are only feasible for a few relatively small supercells (see Table \ref{tab2}), and we must
request that $V_{int}(\{n_{i}\})$ has but a small number of parameters. We construct such a representation based on
physical grounds (rather than trial-and-error) and then validate the results by the quality of the fit.

The structure of polar surfaces is expected to be dominated by electrostatic interactions. For example, for the polar
GaAs (001) surface it was shown that surface energy differences between different orderings are well described by a
simple electrostatic model. \cite{Northrup} We therefore include electrostatic interaction in $V_{int}(\{n_{i}\})$ by
treating surface Cr ions as point charges $q$ interacting via classical Coulomb forces screened by a dielectric
constant $\epsilon$. We started by assuming that the positions of sites A and B do not depend on the environment, but
this simple model was found to be inaccurate. However, it can be significantly improved by including the effect of
atomic relaxations.

For a given configuration $\{n_{i}\}$ the total energy can be reduced by shifting of the surface Cr ions from their
average positions at sites A and B. Such relaxation terms are often important in the thermodynamics of strongly
size-mismatched bulk alloys. Although the Cr lattice sites at the \CHR\ surface are located rather far from each other,
we found that the vertical (normal to the surface) coordinate of a surface Cr ion occupying site A depends rather
strongly on its environment (the shift can be as large as 0.4 \AA, see Appendix \ref{App-A}). Other ions, including Cr atoms at
site B, shift much less, and we therefore only consider relaxations of the A sites. (This approximation is justified by
the resulting high quality of the fitting.) We introduce a vertical coordinate $z_i$ for each occupied A site and
minimize it for the given configuration $\{n_i\}$. (Thus, $z_i$ are treated as adiabatically ``fast'' variables.) The
electrostatic interaction contributes a vertical force depending on the occupation numbers at other sites of the
lattice. The contribution to the total energy depending on $z_i$ is written as
\begin{equation}
H(n,z)=\frac{1}{2}\gamma\sum_{i}\bar{n}_i\left(z_i-z_0\right)^2 - \frac12\sum_{ij}\frac{p_i^2}{\epsilon
d_{ij}^3}\bar{n}_in_j \label{Model1}
\end{equation}
Here we defined $\bar{n}_i=1-n_i$. The first term represents the elastic contribution for each A site as a simple
harmonic oscillator with stiffness $\gamma$ and equilibrium position $z_0$. The second term describes the electrostatic
interaction with B sites. (Small vertical forces from other A sites are neglected.) Here $d_{ij}$ is the distance
between sites $i$ and $j$, and the B sites are assumed to lie at $z=0$. Since $z_i\ll d_{ij}$, we have used the dipole
approximation with $p_i=qz_i$.

Introducing a small parameter $\alpha=q^2/(\gamma\epsilon a^3$), where $a$ is the 2D hexagonal lattice parameter, and
minimizing (\ref{Model1}) with respect to $z_i$, we obtain, to first order in $\alpha$:
\begin{equation}
z_i=z_0+\alpha z_0 R_i \label{Model2}
\end{equation}
where $R_i=\sum_j n_j\zeta_{ij}$  and $\zeta_{ij}=(a/d_{ij})^3$. This relation agrees perfectly with relaxation data for surface supercells (see
Appendix \ref{App-A}). Substituting $z_i$ in (\ref{Model1}), we obtain to first order in $\alpha$:
\begin{equation}
V_{int}(\{n_{i}\})=-\frac{1}{2}V\sum_{ij}\zeta_{ij}n_i\bar{n}_j - X\sum_{i}\bar{n}_iR_{i}^{2} \label{Model3}
\end{equation}
where $V=q^2z_0^2/(\epsilon a^3)$ and $X=\alpha V/2$. The first two-body term represents the
dipolar interactions assuming fixed A site positions $z_i=z_0$. The second three-body term is the lowest-order
correction due to the A site shifts. Note that the parameter $V$ is positive, because an unlike AB bond is longer
than an AA or a BB bond due to the vertical shift, and all Cr ions are positively charged. This transparent physical
mechanism generates an ordering tendency in our system.

The resulting configurational Hamiltonian contains three parameters: $h$, $V$ and $X$. Note that $V_{int}$ vanishes
when all $n_i=0$ or all $n_i=1$, and therefore $h$ gives the positive energy difference between models A and B in Table
\ref{tab1}.

As noted in the Introduction, thin films of \CHR\ demonstrating phase transitions are subject to a tensile epitaxial strain of about 1.5\%. Therefore, in the following we consider two cases: (1) unstrained surface, and (2) surface subject to a 1.5\% in-plane tensile strain.

The three parameters of the configurational model are fitted to the calculated surface energies of several ordered configurations listed in Table
\ref{tab2} and illustrated in Fig.\ \ref{structures}. The standard take-one-out cross-validation (CV) score \cite{Walle} is used to
evaluate the predictive power of the fit. It can be seen that the model (\ref{Model3}) provides an excellent fitting to
the calculated energies for both unstrained and strained surfaces, supporting our assumptions about the physical interaction mechanisms. Note that the parameter $X$ is small compared to $V$ in agreement with our assumptions. Nevertheless, the three-body term is essential for obtaining a good fit (see Appendix \ref{App-B} for further discussion). Without this term a much larger CV score is obtained, and the take-one-out prediction for model B
is about 50 meV off. The quality of the fit is also significantly impaired if the range of the electrostatic interaction is cut off in real space.

One might expect that $V_{int}(\{n_{i}\})$ for nearest neighbors could also have a contribution of non-electrostatic
origin. However, since the surface Cr ions are rather far from each other ($a\approx 5$ \AA) and the surface remains
insulating, this contribution should be short-ranged and relatively small. Indeed, the addition of a nearest-neighbor
pair or three-body (triangle) interaction to $V_{int}(\{n_{i}\})$ did not improve the quality of the fit.

\begin{table}
\caption{Ground state (AFM) and paramagnetic (PM) surface energies for different configurations for the \CHR\ (0001)
surface under zero strain and under 1.5\% tensile epitaxial strain. A$_x$B$_y$ ($d_1\times d_2$) denotes the
configuration with a surface supercell spanned by vectors of length $d_1$ and $d_2$ and containing $x$ ($y$) surface Cr
ions in position A (B). The configurations in the table are shown explicitly in Fig.\ \ref{structures}. The values are
given with respect to the surface energy of model A. Corresponding fitted values of parameters of the configurational
Hamiltonian together with the misfit and the average cross-validation score are also given. Further, the critical
temperature of the ($\sqrt3\times\sqrt3$) to ($1\times 1$) order-disorder transition obtained from MC is also given.
The surface energies, parameters of the configurational Hamiltonian, the misfit and the average cross-validation score
are all given in meV while the critical temperature is in K.}
\begin{tabular}{|l|c|c|c|c|}
\hline
 & \multicolumn{2}{c|}{Unstrained} & \multicolumn{2}{c|}{Strained} \\
\cline{2-5}
 & AFM & PM & AFM & PM \\
\hline
B ($1\times1$) & 168 & 160 & 77 & 91 \\
\hline
AB ($1\times2$) & $-51$ & $-48$ & $-71$ & $-64$ \\
\hline
A$_2$B ($1\times3$) & $-42$ & $-41$ & $-55$ & $-51$\\
\hline
AB$_2$ ($1\times3$) & $-2$ & $-2$ & $-41$ & $-31$ \\
\hline
A$_2$B ($\sqrt{3}\times\sqrt{3}$) & $-62$ & $-57$ & $-70$ & $-65$ \\
\hline
AB$_2$ ($\sqrt{3}\times\sqrt{3}$) & $-28$ & $-22$ & $-60$ & $-49$ \\
\hline
$h$ & 168 & 160 & 76 & 90 \\
\hline
$V$ & 64 & 64 & 52 & 56 \\
\hline
$X$ & 1.6 & 1.4 & 1.2 & 1.0 \\
\hline
misfit & 1 & 1 & 1 & 1 \\
\hline
CV & 2 & 7 & 3 & 3 \\
\hline
$T_C$ & $165\pm5$ & $165\pm5$ & --- & $50\pm10$ \\
\hline
\end{tabular}
\label{tab2}
\end{table}

The main effect of tensile strain on the configurational Hamiltonian is the decrease of the parameter $h$ by about a factor of two compared to the unstrained surface. This effect can be understood by noting that $h$ represents the local preference of the bulk-like surface site A over the interstitial site B. Under tensile strain the lattice expands, leaving more space available for the Cr stom at site B. This reduces the interstitial pressure and thereby the energy cost of occupying site B.

\subsection{Effect of magnetic ordering}

\begin{figure}[htb]
\begin{center}
\includegraphics[width=0.45\textwidth]{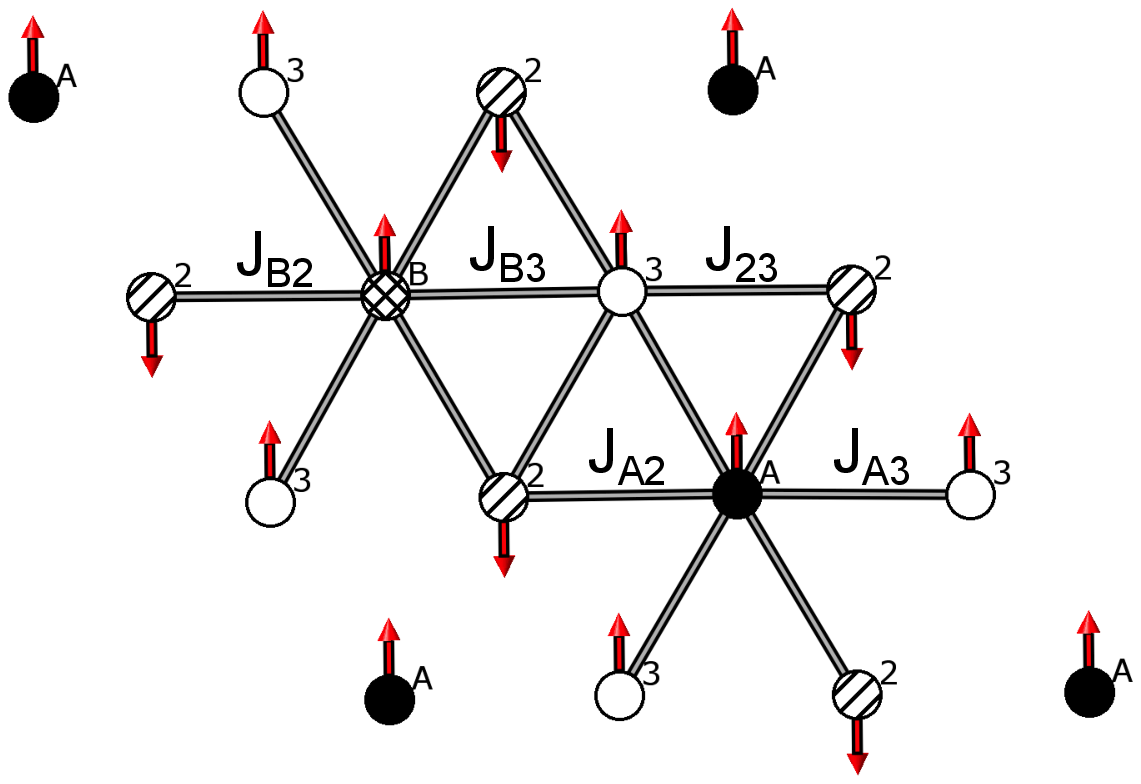}
\end{center}
\caption{Magnetic model for the (0001) \CHR\ surface. We consider three closest to the surface Cr monolayers with four types of Cr ions: site A (filled circle) and site B (hatched circle) from the surface layer, site 2 (striped circle) and site 3 (empty circle) from second and third closest to the surface Cr monolayers, respectively. Nearest-neighbor exchange parameters between different types of Cr ions are denoted by thick gray lines. The red arrows show direction of local magnetic moment in the bulk-like AFM order.}
\label{exchanges}
\end{figure}

So far we have discussed the fitting of the configurational Hamiltonian to surface energies for antiferromagnetically
ordered supercells. The directions of the local moments at B sites were assigned similar to A sites, continuing the
bulk antiferromagnetic structure. The surface magnetic structure may, however, be different from the bulk one. We
checked this by recalculating the surface energies for different magnetic configurations of a few Cr sites closest to
the surface. These sites included A and B sites, as well as the two Cr sites in the underlying buckled honeycomb Cr
layer (types 2 and 3 in the order of depth, see Fig. \ref{exchanges}). Assuming that Cr ions of the same type always have the same spin
direction, for each input surface configuration we calculated the total energy for all possible configurations of the
four near-surface Cr sites (A, B, 2, and 3), while keeping the rest of the slab in its bulk magnetic structure.

We found that the lowest surface energy corresponds to the continuation of the bulk magnetic structure for all surface configurations with the exception of surface model B ($1\times1$). For this model the surface energy is reduced by 43~meV by flipping of the local moment on site 2, thereby making it parallel to those on sites B and 3. Nevertheless, the surface energy of model B listed in the first column of Table \ref{tab1} and used in the fitting corresponds to the continuation of the bulk structure. This preserves consistency with the surface energies of other configurations. The effect of this choice on thermodynamics is small, because surface model B has a large surface energy.

Magnetic disorder present at finite temperatures may affect the relative energies of different surface models and thereby influence the thermodynamic properties. A complete solution requires that the structural ($n_i$) and magnetic degrees of freedom are both included in the effective Hamiltonian. We did not attempt to construct such a Hamiltonian, but rather considered the effect of complete magnetic disorder in the paramagnetic phase on the structural interaction parameters. To this end, for each of the input surface models we have fitted the surface energies to a surface Heisenberg Hamiltonian
\begin{equation}
\mathcal{H}=-\frac{1}{2}\sum_{ij}J_{ij} \mathbf{S}_{i} \cdot \mathbf{S}_{j} -\sum_{i}H_{i}S_{i}^z +N_sE_s^{PM}
\label{Heisenberg}
\end{equation}
Here summation runs over Cr ions belonging to one of the four types defined above (A, B, 2, 3), and $\mathbf{S}_{i}$ is a unit vector parallel to the local moment of the $i$-th ion. Assuming that the exchange coupling does not extend further than in the bulk, the only nonzero exchange parameters are those between nearest-neighbor Cr ions of different types (see Fig.\ \ref{exchanges} for an illustration). In addition, each Cr ion interacts with an effective exchange field $H_{i}$ set up by the bulk. We assumed that Cr ions of the same type are equivalent. Under these assumptions the number of parameters reduces to 10: $H_A$, $H_B$, $H_2$, $H_3$, $J_{A2}$, $J_{A3}$, $J_{B2}$, $J_{B3}$, $J_{23}$, and $E_s^{PM}$. For $1\times1$ surface models only seven of these parameters remain. We have fitted these parameters using 8 magnetic configurations for the $1\times1$ surface models and using 16 magnetic configurations for models with a larger unit cell. Good fits were obtained for all surface models. Table \ref{tab3} lists the fitted parameters and the misfits for surface models A, B, and A$_2$B. These parameters will be discussed further in Section \ref{surf-mag}.

\begin{table}
\caption{Fitted parameters of Eq.\ (\ref{Heisenberg}) and misfits $\Delta$ (all in meV units) for surface models A, B, and A$_2$B. The subscripts of exchange fields $H_i$ and pair parameters $J_{ij}$ refer to the corresponding Cr sites near the surface (see text). Last column: corresponding values in bulk \CHR\ using values from Ref.\ \onlinecite{Siqi}.}
\begin{tabular}{|l|c|c|c|c|}
\hline
         & A & B & A$_2$B  & Bulk \\
\hline
$H_A$    & $0.6$   &         & $0.9$                         & $J_5^b=-2.2$ \\
\hline
$H_B$    &         & $74.9$  & $69.1$                        & $J_5^b=-2.2$ \\
\hline
$H_2$    & $-37.5$ & $-16.8$ & $-29.1$                       & $V^b+3J_4^b=-5.7$ \\
\hline
$H_3$    & $-1.4$  & $4.9$   & $-0.2$                        & $3J_3^b+3J_4^b+J_5^b=10.3$ \\
\hline
$J_{A2}$ & $4.7$   &         & $6.4$                         & $J_3^b=2.1$ \\
\hline
$J_{A3}$ & $11.1$  &         & $10.7$                        & $J_4^b=3.0$ \\
\hline
$J_{B2}$ &         & $12.1$  & $4.8$                         & $J_3^b=2.1$ \\
\hline
$J_{B3}$ &         & $5.4$   & $3.5$                         & $J_4^b=3.0$ \\
\hline
$J_{23}$ & $-9.1$  & $0.6$   & $-8.6$                        & $J_2^b=-11.1$ \\
\hline
$\Delta$ & $0.5$   & $0.1$   & $10^{-4}$                     &  ---  \\
\hline
\end{tabular}
\label{tab3}
\end{table}

The parameter $E_s^{PM}$ represents the surface energy in the paramagnetic state with no spin correlations. These
energies  are listed in Table \ref{tab2} along with the configurational interaction parameters fitted to them. Again the three-body term is essential for obtaining a good fit. Without this term much larger CV score is obtained, and the take-one-out prediction for model B is about 43 meV off. As seen,
the paramagnetic energies and the interaction parameters differ little from their AFM values, indicating that
magnetostructural coupling for the unstrained surface is weak.
However, for the strained surface the parameters of the configurational Hamiltonian depend much stronger on the magnetic state,
indicating substantial magnetostructural coupling. Comparison of different columns of Table \ref{tab2} shows that the
effect of magnetic disorder for the strained surface is qualitatively opposite to that of the tensile strain.

\subsection{Ground-state phase diagram}

\begin{figure}[htb]
\begin{center}
\includegraphics[width=0.45\textwidth]{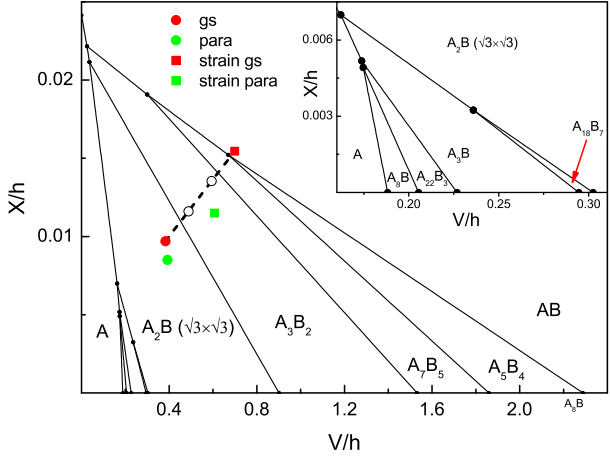}
\end{center}
\caption{Ground state phase diagram for the configurational Hamiltonian. The ground state configurations are shown in
Figure \ref{structures}. Red (green) circles and squares denote the values of parameters of the Hamiltonian fitted to
the ground state (paramagnetic) surface energies for unstrained and strained \CHR\ (0001) surface, respectively. The
dashed line denote a strain path and the open black circles denote the parameters for intermediate strains.}
\label{gsdiagram}
\end{figure}

The search for the likely ground states of our model was performed by a direct enumeration of all configurations for
unit cell sizes up to $6\times 6$. The resulting ground-state phase diagram is shown in Figure \ref{gsdiagram}. The
Hamiltonian fitted to the surface energies of AFM unstrained supercells (first column of Table \ref{tab1}) lies deep
within the region where A$_2$B ($\sqrt3\times\sqrt3$) is the ground state. Due to the long-range character of
electrostatic interactions it is possible that ground states with cell size larger than $6\times 6$ may appears in
certain regions of the parameter space. However, such complicated orderings would be easily destroyed by thermal
fluctuations. We therefore assume that such orderings, even if present, are irrelevant for the thermodynamic properties
at temperatures where equilibration is kinetically achievable; see discussion on kinetic energy barriers below.

\section{Configurational thermodynamics}\label{thermo}

First we have studied the thermodynamics of our model within the mean-field approximation (MFA). We considered ordered
structures A ($1\times1$), AB ($1\times2$), A$_2$B ($\sqrt{3}\times\sqrt{3}$), and A$_3$B$_2$
($\sqrt{3}\times\sqrt{7}$), which appear in the region of the parameter space relevant for \CHR. We did not include
complicated orderings like A$_7$B$_5$ or A$_5$B$_4$, because, as noted above, they are expected to appear only at very
low temperatures. The free energy of each phase was calculated in MFA; the equilibrium phase at a given temperature is
the one with the lowest free energy. (Since the concentration of sites B is not conserved, the equilibrium phase is
always single-phase.)

\begin{figure}[htb]
\begin{center}
\includegraphics[width=0.45\textwidth]{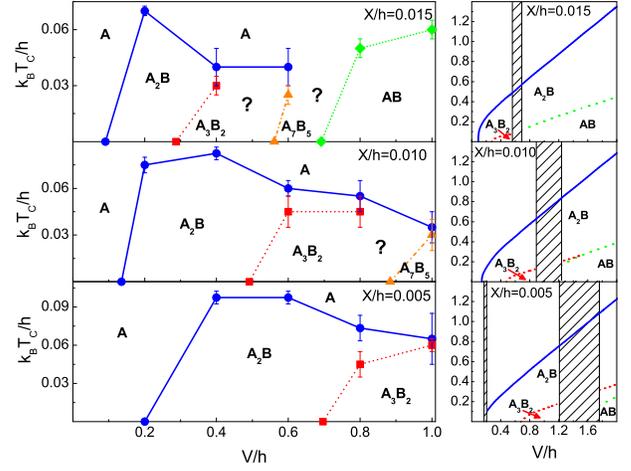}
\end{center}
\caption{Temperature phase diagram for the configurational Hamiltonian obtained by MFA (right) and MC (left). Solid blue,
dashed red, dash-dotted orange, and dotted green lines denote transition temperature to A$_2$B ($\sqrt{3}\times\sqrt{3}$), A$_3$B$_2$,
A$_7$B$_5$, and AB orderings, respectively. In the MC case these lines are guides to the eye. Low-temperature MFA
solutions are not shown in the patterned regions, because the corresponding ground states were not considered in the calculations.}
\label{temppd}
\end{figure}

The MFA results are shown in Fig.\ \ref{temppd} (panels on the right-hand side). At small $V/h$ or $X/h$, where the
ground state is A, the surface never orders and remains A-type at all temperatures.\cite{comment0} Where A$_2$B is the
ground state there is a continuous order-disorder transition from A$_2$B-type to A-type. As the magnitude of $V/h$ or
$X/h$ is increased, the critical temperature (in units of $h$) increases. This trend continues even when in the region
where A$_3$B$_2$ is the ground state. In this region, as temperature increases from zero, the A$_3$B$_2$-type structure
undergoes a first-order transition to A$_2$B-type, which then transforms to A-type at higher temperatures.\cite{Comment1}

In the region where AB is the ground state, the situation depends on $X/h$. For small $X/h$ the AB-type structure
undergoes a series of first-order transitions to A$_3$B$_2$-type and then to A$_2$B-type; the latter then further
transform to A-type. At larger $X/h$ or $V/h$ the A$_3$B$_2$-type ordering disappears and AB transforms directly to
A$_2$B.

Due to strong geometric frustration, the MFA calculations are unreliable and only provide a reference for comparison
with more accurate calculations. We performed Monte Carlo (MC) simulations on $L\times L$ triangular lattices with
periodic boundary conditions. We usually used $L=30$ as this size is commensurate with all the relevant orderings. In
the loop over the lattice sites, a new state with the changed occupation number is tried and accepted or rejected using
the Metropolis algorithm. The evaluation of the energy difference involves an expensive calculation of the long-range
interaction part (\ref{Model3}). For the two-body term this can be done using Fourier transforms. On the other hand,
the direct calculation of the three-body term in Fourier space would be very expensive, because it requires a double
summation over $\mathbf{q}$. Instead, we calculate the Fourier transform of $R_i$ as
$R_{\mathbf{q}}=J_{\mathbf{q}}n_{\mathbf{q}}$, transform it back to real space using a fast Fourier transform (FFT)
technique, and then calculate the three-body term in real space. We thus replace one sum over $\textbf{q}$ by an FFT,
which significantly reduces the computational cost for a large $L$. With this procedure the calculations could be
performed for lattices with up to $L=36$ using a few million (a few hundred thousand) MC steps per site for
accumulating averages (for equilibration). These restrictions were not always sufficient to obtain quantitatively
accurate results (see below), but a qualitative understanding of the phase diagram could be achieved.

The ordering type was identified by analyzing the structure factor $I(\mathbf{q})=\left|n(\mathbf{q})\right|^2$, where
$n(\mathbf{q})$ is the Fourier transform of $n_i$. All phase transitions between different ordered phases that we found
are required by symmetry to be first-order. The order of order-disorder transitions was determined by analyzing the
scaling behavior of the fourth-order energy cumulant. \cite{Challa} The transition temperatures were found from the
peaks of the heat capacity for first-order transitions and from the finite-size scaling behavior of the fourth-order
cumulant of the corresponding order parameter \cite{Binder} for continuous transitions. For scaling analysis we used
lattices with $L=30$, 33, 36 for A$_2$B-type ordering and $L=30$, 32, 34, 36 for AB-type ordering.

The results of MC simulations are shown in Fig.\ \ref{temppd} (panels on the left-hand side). Similarly to MFA, for
small values of $V/h$ and $X/h$ where A$_2$B is the ground state an order-disorder transition to the A-type phase is
observed. Our procedure identifies this transition as being everywhere continuous (second-order), except perhaps for
small values of $V/h$, where the results suggest the proximity of a first-order transition. Note that the existence
of a tricritical point was reported for the phase diagram of a related 2D hexagonal Ising model with AFM nearest
neighbor and FM second-neighbor interactions. \cite{Mihura}

As expected due to strong geometric frustration, the critical temperature of the A$_2$B-type ordering transition is
strongly suppressed compared to MFA. For the parameters corresponding to magnetically ordered unstrained \CHR\ surface
we found $T_C=165\pm5 K$ in MC compared to 600~K in MFA. If $V/h$ or $X/h$ are increased, initially $T_C$ also
increases due to the stabilization of the A$_2$B-type structure relative to A-type. However, the increase of $V/h$ or
$X/h$ also leads to stronger frustration, which tends to decrease $T_C$. This competition results in a maximum of $T_C$
as a function of these parameters. Note that the latter effect is absent in MFA (which is insensitive to frustration),
which thereby completely fails for large $V/h$ or $X/h$, incorrectly predicting that $T_C$ should keep increasing.

At low temperatures the system becomes difficult to equilibrate, and the equilibration time increases with increasing
$V/h$ or $X/h$. This is likely associated with increased frustration. In particular, we were unable to equilibrate
the system at low temperatures in the parameter range where the ground state is different from A and A$_2$B, and the
system remained in the initially chosen ordering state. In this case we chose the initial state to be the ground state
structure for the given set of parameters. \cite{Comment2}  In particular, we performed MC simulations by increasing
the temperature starting from the A$_3$B$_2$, A$_7$B$_5$, and AB ground states. Usually these structures underwent a
first-order transition to A$_2$B-type phase, which transforms to A-type under further heating (as discussed above).
These two transitions are often very close to each other. Unfortunately, the temperatures of both first-order and
second-order transitions could usually be determined only with fairly large error bars. For first-order transitions
these error bars are due to hysteretic behavior, while for the second-order transition they result from strong
fluctuations and limited averaging time. Often the error bars for these two transitions overlapped, indicating that the
appearance of the intermediate A$_2$B phase might be spurious, and that the ground state structure may in reality
transform directly to A-type. This is exactly what happens for large values of $V/h$ and $X/h$ when AB is the ground
state.

\section{Comparison with experiment} \label{sec-exper}

Using the parameters fitted to the AFM surface energies for \CHR\ surface the temperature dependence of the fraction of
surface Cr ions occupying sites B was found from MC simulations, see Fig.\ \ref{magT}. At low temperatures, when the
surface is A$_2$B-type, the concentration is close to the ideal value of 1/3 for the ground-state A$_2$B structure. As
the temperature increases there is an order-disorder transition at $T_C\approx165 K$. In a temperature region around
the transition the B-site fraction increases to about $40\%$ and then stays approximately constant for temperatures
well above room temperature. This result is in reasonable agreement with the room-temperature fraction of $\sim 33\%$
found from SXRD. \cite{Gloege}

We found that the heat capacity has a broad shoulder above the critical temperature, indicating the persistence of
strong short-range order well above room temperature. This is a direct consequence of geometric
frustration.\cite{Comment3}

The $1\times1\to\sqrt{3}\times\sqrt{3}$ ordering transition found above can be identified with the high-temperature
phase transition observed in LEED.\cite{Bender} $T_C\approx165$~K produced by MC simulations agrees with the
observed\cite{Bender} $T_C\sim150$ K. However, a second phase transition back to $1\times1$ at about 100~K was also
observed in these LEED measurements.\cite{Bender} This transition does not appear in our calculations for the
unstrained surface. It was suggested \cite{Bender,Gloege} that this second transition at 100 K may be induced by
magnetostructural coupling. As explained above, our calculations do not support this hypothesis for an unstrained
surface. Our theory predicts that the surface of an unstrained \CHR\ crystal should undergo only one
$1\times1\to\sqrt{3}\times\sqrt{3}$ ordering transition.

\begin{figure}[htb]
\begin{center}
\includegraphics[width=0.45\textwidth]{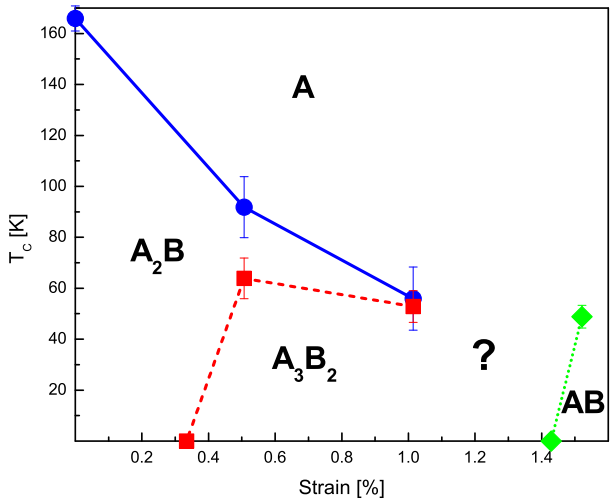}
\end{center}
\caption{MC transition temperatures for different orderings as a function of tensile strain. Blue circles, red square and green rhombi
denote transition temperatures below which A$_2$B ($\sqrt{3}\times\sqrt{3}$), A$_3$B$_2$
($\sqrt{3}\times\sqrt{7}$), and AB ($1\times2$), orderings respectively set in. Lines connecting the points are guides
to the eye.}
\label{strain}
\end{figure}

For the analysis of the trend introduced by the strain we consider a continuous path in the parameter space assuming
that all the parameters change linearly with strain, interpolating between the AFM surface energies found for 0\% and
1.5\% strain. As shown in Fig.\ \ref{gsdiagram}, the strain changes the ground state ordering from A$_2$B to AB,
passing through A$_3$B$_2$ and A$_7$B$_5$ in between. The effect on structural thermodynamics is illustrated in Fig.\
\ref{strain}, where the temperatures of different phase transitions are shown as a function of strain. The A$_2$B-type
ordering temperature is decreased by strain. At a certain value of strain the ground state changes to A$_3$B$_2$. Under
heating this structure transforms to A$_2$B, which then disorders at a higher temperature. For larger strains, however,
the A$_2$B phase disappears, and A$_3$B$_2$ transforms directly to A ($1\times1$). As the strain further increases, the
A$_7$B$_5$ phase is expected to appear at low temperatures (not shown in Fig.\ \ref{strain}), and at yet a larger
strain the AB phase sets in.

The phase transitions at low temperatures may be unobservable for kinetic reasons. We have calculated the activation
energy $E_b$ for the jumping of a Cr ion from site A to B. Smooth barrier profiles were obtained with $E_b$ equal to
$0.4$ eV and $0.3$ eV for free and 1.5\% strained surfaces, respectively. The frequency of thermally activated jumps
between sites A and B can be then estimated as $\gamma \sim \gamma_0 e^{-\frac{E_b}{k_BT}}$ where $\gamma_0$ is the
attempt frequency on the order of a typical phonon frequency $\sim10^{13}$ s$^{-1}$. (Or perhaps an order of magnitude
smaller for thermal phonons at low $T$.) It follows that at room temperature the typical hopping time is of the order
of $10^{-8}$~s. On the other hand, the blocking temperature below which the kinetics is frozen is about 100~K.
Therefore, the equilibrium phase transformations predicted for temperatures notably below 100 K are unobservable, and
the system is expected to be trapped in the structural state corresponding to equilibrium near the blocking
temperature.

Our results support the hypothesis \cite{Bender,Gloege} that magnetostructural coupling plays an important role in the
origin of the two phase transitions observed in LEED for a thin strained film. \cite{Bender} The following picture can
be suggested. At low temperatures the parameters of the configurational Hamiltonian correspond to the AFM-ordered
strained surface. As seen in Fig.\ \ref{strain}, in this case the equilibrium state near the blocking temperature is
disordered and has a $1\times1$ symmetry. Higher temperatures introduce partial spin disorder which, as discussed
above, changes the parameters of the Hamiltonian similarly to a decrease of strain. This leads to the enhancement of
the A$_2$B-type ordering temperature (Fig.\ \ref{strain}), which becomes higher than the blocking temperature and then
overtakes the temperature of the system. In this picture this point corresponds to the low-temperature transition
observed in LEED. As the temperature further increases, the system passes through the conventional disordering
transition.

The above scenario requires the surface to be under an exactly right amount of strain, and it implies that the phase
transitions are very sensitive to the growth conditions. This indirectly agrees with the fact that no phase transitions
were observed for a thicker \CHR\ film. \cite{Takano} In this case the parameters of the Hamiltonian at low
temperatures may correspond to a larger strain, which keeps the system disordered at all temperatures.

The main drawback of the proposed mechanism of the reentrant phase transition is that magnetic disorder is assumed to
influence the structural energetics at temperatures that are significantly below the N\'eel temperature. Note, however,
that the effect of magnetic disorder is to a notable extent mediated by the reduction of the parameter $X$, which
describes the relaxation stiffness of site A relative to the electrostatic forces. But since the exchange coupling of
site A to the bulk is quite weak (see Table \ref{tab3}), the magnetic disorder affects this site already at low
temperatures (see Section \ref{surf-mag}). This factor gives some support to the proposed mechanism. For a more
detailed consideration the magnetic degrees of freedom would have to be included in the Hamiltonian. We did not attempt
this due to the limited amount of experimental information on the atomic structure of the surface.

\section{Surface electronic structure}\label{el-str}

\begin{figure}[htb]
\begin{center}
\includegraphics[width=0.45\textwidth]{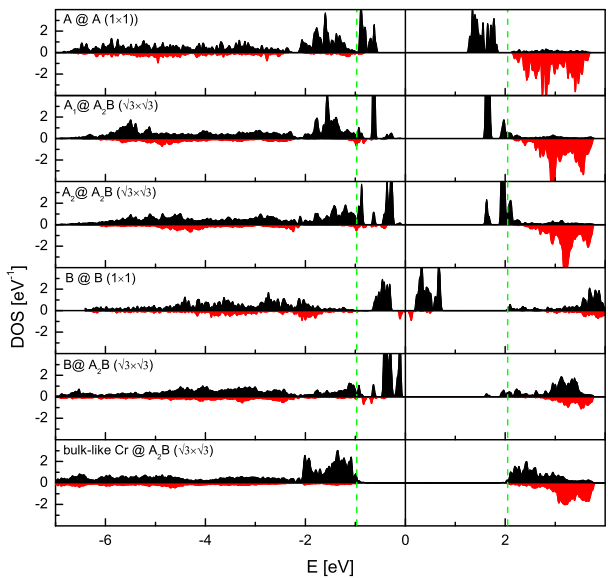}
\end{center}
\caption{Spin resolved densities of states (DOSs) for A and B surface Cr ions and bulk-like Cr ion in the middle of the slab calculated for surface models: A ($1\times1$), B ($1\times1$), and A$_2$B ($\sqrt3\times\sqrt3$). Two nonequivalent A surface Cr ions for the A$_2$B ($\sqrt3\times\sqrt3$) surface model (see the text) are denoted by A$_1$ and A$_2$. Majority and minority DOSs are plotted on positive and negative y axis, respectively. Energy zero is set to the valence band maximum in the A$_2$B ($\sqrt3\times\sqrt3$) surface model. DOSs obtained for different slabs are aligned by semicore O $2s$ states for bulk-like oxygen ions in the middle of the slabs. The green dashed vertical lines denote the bulk band gap. The DOS within the bulk band gap comes from the surface states. }
\label{dos}
\end{figure}

In this section we discuss the electronic structure of the \CHR\ (0001) surface. Fig.\ \ref{dos} shows partial
densities of states (DOS) for A and B surface Cr ions for A, B, and A$_2$B surface models. For comparison we include
the partial DOS for the Cr ion in the middle of the A$_2$B slab, which is similar to bulk \CHR. \cite{Siqi}  The DOS
plots for different supercells are aligned using the semicore $2s$ states for bulk-like oxygen ions in the middle of
the slabs.

For model A$_2$B there are two A-site Cr ions in the surface supercell, which are denoted as A$_1$ and A$_2$. The
A$_2$B ordering makes these sites inequivalent due to the directional character of the bonds, which was discussed in
Section \ref{config}. The partial DOS for A$_1$ and A$_2$ sites are similar, except for a shift of about 0.3~eV. This
electrostatic shift is due to the fact that site A$_1$ is further away from the O sites in the subsurface layer than
A$_2$.

The partial DOS for sites A in $1\times1$ and $\sqrt3\times\sqrt3$ surface supercells are qualitatively similar up to a moderate upward shift in the $\sqrt3\times\sqrt3$ supercell. The same can be said for site B, but the shift is in the opposite direction.
The partial DOS for site A shows surface states in the bulk band gap both close to the valence band maximum and to the
conduction band minimum. Site B introduces surface states originating from the valence band but extending deeper into
the bulk band gap. In both cases there is strong hybridization with the subsurface O ions. Partial DOS for deeper layers (not shown) shows that the surface states decay within 3-4 Cr monolayers from the surface.

\section{Surface magnetism} \label{surf-mag}

In Section \ref{sec-exper} we have seen that magnetic disorder may affect the surface phase transitions through a
peculiar magnetostructural effect. On the other hand, the equilibrium magnetization of the \CHR\ (0001) surface enables
interesting spintronic applications. \cite{XiHe,Andreev,Kirill,NingWu}  For these reasons it is interesting to consider
the magnetic properties of the \CHR\ (0001) surface at finite temperatures.

The surface Heisenberg Hamiltonians were obtained in Section \ref{surf-struc} for different surface models (see Table
\ref{tab3} for the parameters for models A, B, and A$_2$B). A common feature for all surface models is very strong exchange coupling of site B and weak coupling of site A to the bulk. This is expected, because all of the four bulk-like nearest and next-nearest neighbors of site A are absent; in spite of its large vertical relaxation, the remaining couplings do not compensate for this. The corresponding parameters for models A and A$_2$B are quite similar. Although there are some differences for models B and A$_2$B, configurations close to model B are statistically rare due to the fact that the equilibrium concentration of sites B is approximately 1/3. The fitted parameters differ significantly from the bulk couplings, which is a result of large ionic relaxations near the polar surface. (The parameters calculated as if the bulk exchange parameters \cite{Siqi} do not change near the surface are listed in the last column of Table \ref{tab3}.)

\begin{figure}[htb]
\begin{center}
\includegraphics[width=0.45\textwidth]{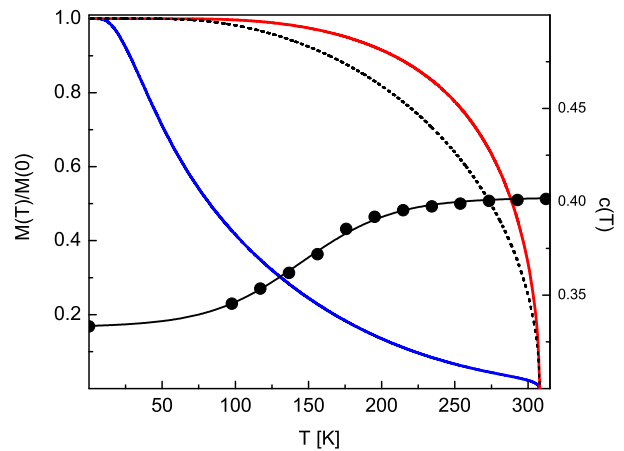}
\end{center}
\caption{The temperature dependence of magnetizations of surface sites $M(T)$ for the A$_2$B ($\sqrt{3}\times\sqrt{3}$) surface model. Solid blue and red lines denote magnetization of site A and B, respectively. Dotted black line denote bulk sublattice magnetization. Black circles show MC results for the temperature dependence of the concentration of surface Cr ions occupying site B. The solid black lines is the best fit to MC data.}
\label{magT}
\end{figure}

To calculate the temperature dependence of magnetizations for surface sites A and B, we used the mean-field
approximation applied to the quantum spin-$3/2$ version of the Heisenberg model (\ref{Heisenberg}). We considered the
A$_2$B surface model, since it is predicted to be the ground state for the unstrained \CHR\ surface. Since the
magnetostructural coupling is weak, we expect that the surface site magnetizations are largely independent on the
surface structure. We assumed that the exchange fields in (\ref{Heisenberg}) are proportional to the bulk mean-field
sublattice magnetization normalized to the experimental N\'eel temperature. The resulting MFA surface site
magnetizations are shown in Fig.\ \ref{magT}.

Since site B is strongly exchange-coupled to the bulk (Table \ref{tab3}), its magnetization largely follows the bulk
sublattice magnetization. On the other hand, site A is weakly coupled to the bulk. As a result, its magnetization is
substantially reduced and exhibits an inflection point. This inflection could be observed in the temperature dependence
of surface magnetic response in any experiment sensitive to the surface magnetism. Note that at 100 K site A is
predicted to have already lost about 60\% of its magnetization at $T=0$. As mentioned above in Section
\ref{sec-exper}, this behavior lends some support to the magnetostructural coupling mechanism of the reentrant
structural phase transition observed in Ref.\ \onlinecite{Bender}.

\section{Conclusions}\label{summary}

Based on first-principles total energy calculations and Monte Carlo simulations, we proposed a detailed microscopic
model explaining the mechanisms of phase transitions at the stoichiometric \CHR\ (0001) surface. Partial occupation of
two surface Cr sites gives rise to complicated thermodynamic properties. Interaction is dominated by electrostatic
forces, which promote ordering, and contains a smaller but still important contribution from atomic relaxations. The
ground state is ordered with a $\sqrt{3}\times\sqrt{3}$ unit cell; it undergoes a continuous order-disorder transition
at $T_C\approx165$ K. Tensile epitaxial strain has a strong effect on the surface energetics, enhancing frustration,
introducing new ground states and additional phase transitions. Magnetostructural coupling also plays an important role
in the structural thermodynamics of the strained surface. Based on these results, we proposed an explanation of the
reentrant $1\times1\to\sqrt3\times\sqrt3\to1\times1$ phase transitions observed experimentally on thin \CHR\ (0001)
films grown on Cr. \cite{Bender}

\begin{acknowledgments}

We are grateful to P.\ A.\ Dowben for useful discussions. This work was supported by NSF through Nebraska MRSEC (Grant
No.\ DMR-0820521) and Nebraska EPSCoR (Grant No.\ EPS-1010674). K.\ D.\ B.\ is also supported by the Research
Corporation through a Cottrell Scholar award.

\end{acknowledgments}

\appendix

\section{Surface relaxations} \label{App-A}

Here we include the data on the atomic relaxations at the \CHR\ (0001) surface and provide a justification for model
(\ref{Model1})-(\ref{Model2}). Table \ref{tab4} lists the interlayer distances for A ($1\times1$) and B ($1\times1$)
surface models. Strong inward relaxations are observed, as expected for a polar surface. For model A ($1\times1$)
the relaxations extend up to the fifth atomic layer, while for model B ($1\times1$) they propagate much further, because the occupation
of the interstitial site B introduces a stronger disturbance. Interlayer relaxations for model A ($1\times1$) are in reasonable agreement with other theoretical calculations. \cite{Rohr,Rehbein,Rohrbach} Although there are notable deviations from the LEED data, \cite{Rohr} we need to remember that the latter correspond to the actual surface termination but were fitted assuming the A ($1\times1$) model.

\begin{table}[ht]
\caption{Surface interlayer distances in $\%$ of the bulk interlayer distances for A ($1\times1$) and B ($1\times1$)
surface models. The bulk interlayer Cr-O and Cr-Cr distances are $0.94$ {\AA} and $0.39$ {\AA}, respectively
\cite{Siqi}. Here A($n$) denote $n$th atomic layer from the surface which has ions of type A. Our results are compared with existing literature. Here HF and MD denote Hartree-Fock and Molecular Dynamics methods, respectively. The experimental data (Exp) were obtained using LEED.}
\begin{tabular}{|c|c|c|c|c|c|c|}
\hline
                & \multicolumn{5}{|c|}{A ($1\times1$)}                                                            & B ($1\times1$) \\
\cline{2-7}
                & LSDA+U  & GGA+U\footnotemark[1] & HF\footnotemark[2] & MD\footnotemark[3] & Exp\footnotemark[3] & LSDA+U \\
\hline
Cr($1$)-O($2$)  & $-56.4$ & $-60$                 & $-50$              & $-58$              & $-38$               & $-179.7$ \\
\hline
O($2$)-Cr($3$)  & $+7.3$  & $+12$                 & $+3.3$             & $0$                & $-21$               & $-9.2$   \\
\hline
Cr($3$)-Cr($4$) & $-41.4$ & $-44$                 & $0$                & $-36$              & $-25$               & $-37.9$  \\
\hline
Cr($4$)-O($5$)  & $+10.8$ & $+9.2$                & $0$                & $+17$              & $+11$               & $+18.5$  \\
\hline
O($5$)-Cr($6$)  & $+0.8$  &                       &                    &                    &                     & $+16.0$  \\
\hline
Cr($6$)-Cr($7$) & $-2.4$  &                       &                    &                    &                     & $-45.4$  \\
\hline
Cr($7$)-O($8$)  & $+0.7$  &                       &                    &                    &                     & $10.6$   \\
\hline
\end{tabular}
\footnotetext[1]{Ref. \onlinecite{Rohrbach}}
\footnotetext[2]{Ref. \onlinecite{Rehbein}}
\footnotetext[3]{Ref. \onlinecite{Rohr}}
\label{tab4}
\end{table}

\begin{figure}[htb]
\begin{center}
\includegraphics[width=0.45\textwidth]{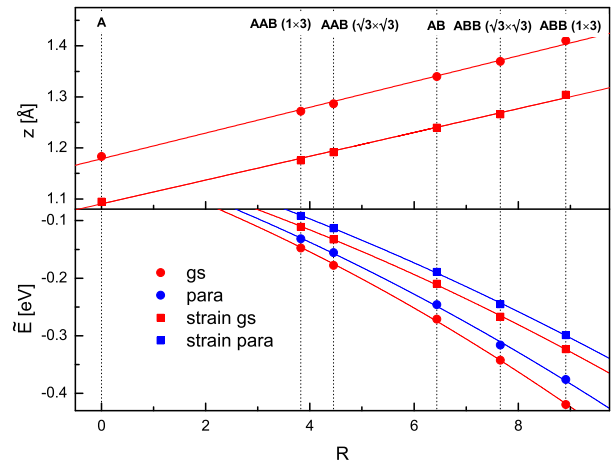}
\end{center}
\caption{Upper panel: Vertical coordinate $z$ of a Cr ion at site A for different surface models as a function of $R_i$
defined after Eq.\ (\ref{Model2}). Circles (squares) correspond to the unstrained (strained) surface. Solid lines are linear fits to the data.
Lower panel: $\widetilde E$ as a function of $R$ (see Eq.\ (\ref{Ehat})) for different surface models. Red (light) symbols correspond to the ground state, and blue (dark) symbols to the paramagnetic state. Circles (squares): data for unstrained (strained) surface. Solid lines are fits to a quadratic function with zero constant term, see Eq.\ (\ref{Ehat}). From bottom to top, the curves are shifted upward by 0, 0.01, 0.01 and 0.03 eV, respectively.}
\label{relax}
\end{figure}

The upper panel of Fig.\ \ref{relax} shows the vertical coordinate of surface Cr ions occupying site A for different
surface models as a function of $R_i$ defined after Eq.\ (\ref{Model2}). This coordinate is referenced with respect to that of the Cr ions occupying sites B averaged over different surface models. (The subsurface O layer was used as an anchor for measuring the $z$ coordinate.) For surface supercells with two inequivalent A-site Cr ions their vertical coordinates were similar, and we used their average. One can see that the calculated data agree very well with Eq.\ (\ref{Model2}) for both unstrained and strained surfaces.

\section{Quality of the fit} \label{App-B}

Here we demonstrate the quality of the fit of \emph{ab initio} energies to the configurational Hamiltonian and explain the importance of the three-body term in Eq.\ (\ref{Model3}).

Note that for all surface models for which \emph{ab initio} energies were calculated, the A sites are equivalent (ignoring the directionality of the bonds on the actual surface, see Section \ref{config}). In this case, the surface energy from Eq.\ (\ref{Model3}) can be rewritten as
\begin{equation}
E=hc-\frac{V}{2}(1-c)(R+\alpha R^2)+\mathrm{const}
\end{equation}
where $R$ is the value of $R_i$ for the A sites. Setting $h=E_B-E_A$, where $E_A$ and $E_B$ are the surface energies for models A ($1\times1$) and B ($1\times1$), we can define
\begin{equation}
\widetilde E \equiv \frac{E-E_A-c(E_B-E_A)}{1-c}=-\frac{V}{2}R-\alpha\frac{V}{2}R^2
\label{Ehat}
\end{equation}
In the lower panel of Fig.\ \ref{relax} we plotted $\widetilde E$ as a function of R using the \emph{ab initio} energies for all considered surface models. We included the data for both strained and unstrained surfaces using both ground state and paramagnetic surface energies. The resulting plots are very well fitted by the quadratic function with a zero constant term, demonstrating the high fidelity of the fit.

The value of the parameter $\alpha$ extracted from the fit ranges from $0.04$ to $0.05$, which, as expected, is a small number. However, the relative importance of the three-body term compared to the two-body term is $\alpha R$. Since for the considered surface models $R$ varies between $4$ and $9$, the relative importance of the three-body term is substantial and reaches 50\%. In the diagrammatic cluster-expansion language one can say that although $\alpha$ is small, the number of corresponding diagrams is large.


\begin{thebibliography}{9}
\bibitem{comment0} Here and below notation like ``A-type'' is used for the partially disordered phase with the ordering pattern of ground state A.
\bibitem{Comment1} The shaded areas in Fig.\ \ref{temppd} denote regions where the ground-state
structure was not included in MFA. Typically these structures have complicated orderings and should undergo first-order
transitions to simpler orderings like A$_3$B$_2$ or AB at fairly low temperatures. Therefore, with exception of the
small rectangle on the left for $X/h=0.005$, the high-temperature behavior in the shaded areas is correct within MFA.
\bibitem{Comment2} Note that due to our inability to equilibrate the system at low temperatures we cannot observe multiple first-order transitions that were predicted to occur in MFA.
\end{thebibliography}
\end{document}